\documentclass[aps,pra,twocolumn,10pt,amsmath,amssymb,showpacs,superscriptaddress,longbibliography]{revtex4-1}%mine

\usepackage{graphicx}% Include figure files
\usepackage{dcolumn}% Align table columns on decimal point
\usepackage{bm}% bold math

\usepackage{psfrag}%not using?
\usepackage[finalnew]{trackchanges}
\newcommand{\myeta}{\xi}

\begin{document}

\title{Impedance matching of inverted conductors: Two-dimensional beam splitters with divergent gain}

\author{Matthew Mecklenburg}
\email{mecklenburg@ucla.edu}
\affiliation{Department of Physics and Astronomy, University of California, Los Angeles, California, 90095}
\affiliation{California NanoSystems Institute, University of California, Los Angeles, California, 90095}

\author{B.C. Regan}
\email{regan@physics.ucla.edu}
\affiliation{Department of Physics and Astronomy, University of California, Los Angeles, California, 90095}
\affiliation{California NanoSystems Institute, University of California, Los Angeles, California, 90095}

\date{\today}% It is always \today, today,
             %  but any date may be explicitly specified February 9, 2010

\begin{abstract}
A thin conducting sheet --- graphene, for example --- transmits, absorbs, and reflects radiation. A sheet that is very thin, even vanishingly so, can still produce 50\% absorption at normal incidence if it has conductivity corresponding to half the impedance of free space.   We find that, regardless of the sheet conductivity, there exists a combination of polarization and angle of incidence that achieves this impedance half-matching condition.  If the conducting medium can be inverted, the conductivity is formally negative and the sheet amplifies the incident radiation.  To the extent that a negative half-match in a thin sheet can be maintained, enormous single-pass gain in both transmission and reflection is possible. Known semiconductors (e.g., gallium nitride) have the optical properties necessary to give large amplification in a structure that is, remarkably, both thin and nonresonant.

\end{abstract}

%42.79.Fm	Reflectors, beam splitters, and deflectors
%42.55.Ah	General laser theory
%78.20.-e	Optical properties of bulk materials and thin films
%Population inversion, 32.80.Xx, 33.80.Be, 42.50.-p

\maketitle
\add{\section{Introduction}}

An electromagnetic medium can be characterized by its permittivity $\varepsilon$ and permeability $\mu$, or equivalently by its index of refraction $n$ and impedance $\eta$ \cite{1995Heald}.  Metamaterials exhibiting the unusual case of a negative index can be engineered  \cite{1968Veselago} and show promise for previously unimagined applications such as perfect lenses  \cite{2000Pendry}.  Negative impedances are analogous to negative indices: while not the usual case, they are not forbidden.  Materials with this property are prepared by creating a population inversion, and are widely used in lasers and other optical amplifiers  \cite{1995Heald}.  This communication describes how a negative-conductivity, properly impedance-matched thin sheet can form an amplifying beamsplitter with enormous single-pass gain.

Much like pellicle beamsplitters, these systems are thin films that divide an incident beam into transmitted and reflected components. The case of reflective amplification is particularly noteworthy, for, in the limit where the film becomes vanishingly thin, the reflection and amplification occur simultaneously upon incidence at the gain medium interface.  This situation stands in marked contrast to other types of optical amplifiers, even those where reflection plays a conspicuous role in enhancing the gain.  For instance,  in disk lasers, also known as \emph{active mirrors}, the gain can be attributed to one medium and the reflection to another  \cite{1981Abate}.  Likewise in fibers with active cladding the gain is most easily viewed as occurring as the evanescent wave propagates in the gain medium  \cite{1966Koester}.  Here, in the thin film limit, the gain and reflection cannot be conceptually decoupled. Thus this classical example identifies a direct connection between reflection and stimulated emission.

\begin{figure}\begin{center}
	\includegraphics[width=0.95\columnwidth]{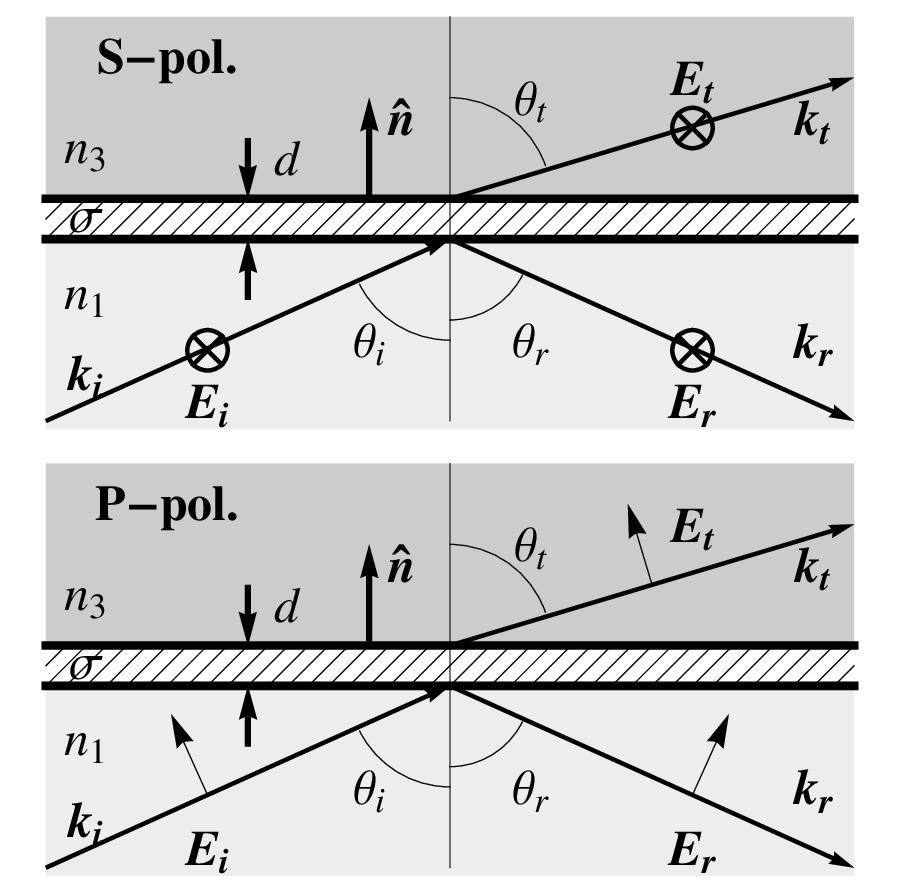}
	\caption{\label{fig:slabgeometry} Diagrams establishing sign conventions for $S$-polarized and $P$-polarized plane waves incident on a conductor sandwiched between media with refractive indices $n_1$ and $n_3$ respectively.  An $\mathbf E$-field directed into the page is indicated with $\otimes$, and the magnetic fields $\mathbf B$ not shown are oriented such that $\mathbf E \times \mathbf B$ is along the wavevector $\mathbf k$.  We consider two cases: the conductor is a sheet with thickness $d= 0$ and two-dimensional conductivity $\sigma_\text{2D}$, or  it is a slab with $d\ne 0$, permittivity $\hat{\varepsilon}_2$, and  three-dimensional conductivity $\sigma_\text{3D}$.}
\end{center}\end{figure}

\begin{figure*}\begin{center}
	\includegraphics[width=0.95\textwidth]{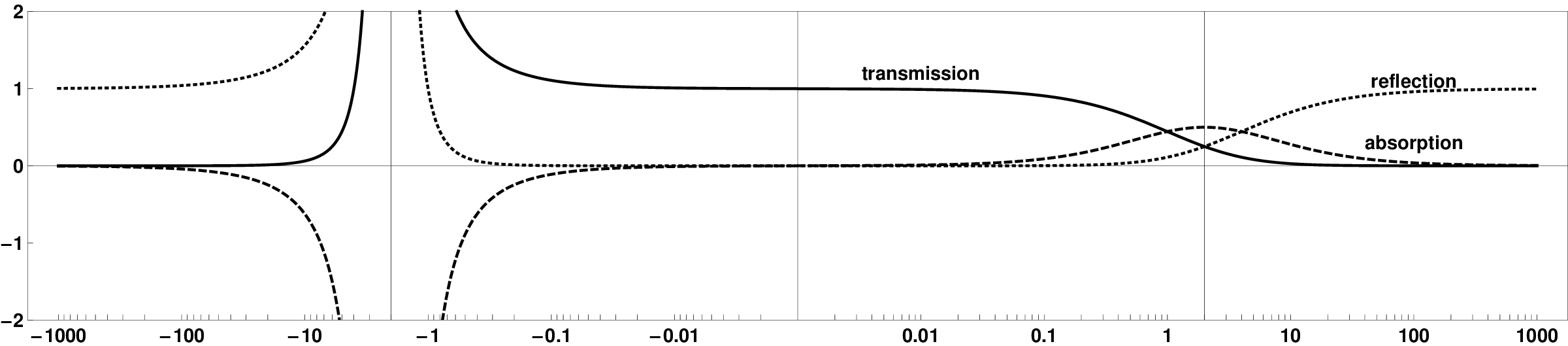}
	\caption{\label{fig:thinsheet} The transmission $T$, reflection $R$, and absorption $A$ by a two-dimensional (2D) conducting plane as a function of $\myeta$, where $\myeta = \sigma_\text{2D}Z_0/n\cos \theta$ for $S$-polarization and $\myeta = \sigma_\text{2D}Z_0 \cos \theta/n$ for $P$-polarization. For a given value of the conductivity $\sigma_\text{2D}$ any magnitude of $\myeta$ can be arranged by judicious choice of the polarization and the angle of incidence $\theta$.  Thus strong coupling ($\myeta=\pm 2$), which gives 50\% absorption for positive conductivity and divergent gain with negative conductivity, is always possible.}
\end{center}\end{figure*}

\section{Sheet}

The reflection, transmission, and absorption of classical electromagnetic waves by a planar conductor is depicted in Fig.~\ref{fig:slabgeometry}.  We begin by treating the case of a conducting sheet of infinitesimal thickness, an algebraically simple but conceptually rich example which captures our  main results. 
We imagine plane waves of angular frequency $\omega$ and wavevector $\mathbf k$ incident on a sheet at an angle $\theta$ relative to the sheet normal vector $\hat{\mathbf{n}}$ (Fig.~\ref{fig:slabgeometry}).   Ohm's law $\mathbf{K}=\sigma_{\text{2D}} \mathbf{E}$ describes the coupling between the surface current density $\mathbf{K}$ and the electric field $\mathbf{E}$.   The two dimensional conductivity $\sigma_{\text{2D}}$ has the dimensionality of conductance, e.g., $\Omega^{-1}$ or $\square/\Omega$.  The third and fourth Maxwell equations give the non-redundant relations that follow from matching the fields at the boundary,
\begin{subequations}\label{eq:boundaryconditions}
\begin{align}
\mathbf{\hat{n}}\times (\mathbf{E}_t-\mathbf{E}_i-\mathbf{E}_r)&=0\qquad \text{and}\\
\mathbf{\hat{n}}\times \lbrack \sqrt{\varepsilon_3/\mu_3}\,\mathbf{k}_t\times \mathbf{E}_t-\qquad\qquad\nonumber\\
\sqrt{\varepsilon_1/\mu_1}(\mathbf{k}_i\times \mathbf{E}_i+\mathbf{k}_r\times\mathbf{E}_r) \rbrack &=(4\pi/c)\mathbf{K},
\end{align}\end{subequations}
where we have used $\mathbf{H}=\sqrt{\varepsilon/\mu}\,\hat{\mathbf{k}}\times \mathbf{E}$.   For simplicity we take all of the materials in the problem to be non-magnetic, with permeability $\mu=1$.  \remove{Since the fields are continuous, the sheet current $\mathbf{K}$ can be considered to be generated by the total electric field on either side of the boundary, or the average of the two.}  Solving these equations with the sign conventions shown in Fig.~\ref{fig:slabgeometry} gives
\begin{subequations}\label{eq:sheetFresnel}
\begin{align}
t_S&=\frac{2 n_1 \cos\theta_1}{n_1 \cos\theta_1 + n_3 \cos\theta_3 +\sigma_\text{2D}Z_0}\\
r_S&=\frac{n_1 \cos \theta_1 - n_3 \cos \theta_3 -\sigma_\text{2D}Z_0}{n_1 \cos\theta_1 + n_3 \cos\theta_3 +\sigma_\text{2D}Z_0}\label{eq:sheet_rs}\\
t_P&=\frac{2 n_1 \cos\theta_1}{\phantom{-}n_1 \cos\theta_3 + n_3 \cos\theta_1+ \sigma_\text{2D}Z_0 \cos \theta_1 \cos\theta_3}\\
r_P&=\frac{-n_1 \cos \theta_3 + n_3 \cos\theta_1 + \sigma_\text{2D}Z_0 \cos \theta_1 \cos \theta_3}{\phantom{-}n_1 \cos\theta_3 + n_3 \cos\theta_1+ \sigma_\text{2D}Z_0 \cos \theta_1 \cos\theta_3}
\end{align}
\end{subequations}
where $r_{S/P}$ and $t_{S/P}$ refer to the normalized ratios of the reflected and transmitted electric fields for the two polarizations $S$ and $P$ ($\mathbf{E}$ $\perp$ and $||$  the plane of incidence respectively). Here $Z_0= 4 \pi/c \simeq 377~ \Omega$ is the impedance of free space, and for the dissipationless, semi-infinite boundary media $n=\sqrt{\varepsilon}$.  The angles of incidence, reflection, and transmission have been re-written as $\theta_i=\theta_r\equiv\theta_1$ and $\theta_t \equiv \theta_3$, where $ \theta_1$ and  $\theta_3$ are related by Snell's law, $n_1 \sin \theta_1=n_3\sin\theta_3$. When the sheet has zero conductivity ($\sigma_\text{2D}=0$), the expressions (\ref{eq:sheetFresnel}) properly reduce to the usual Fresnel equations describing transmission and reflection at the interface between two dielectrics.

Energy conservation gives  $R+T+A=1$, which  relates the absorption ($A$) to the reflection ($R=r^2$) and transmission ($T= t^2 n_3\cos\theta_3/n_1\cos\theta_1$) power ratios.  For the case where the medium is the same on both sides of the sheet, $n_1=n_3\equiv n$ and $\theta_1=\theta_3\equiv\theta$. The results for both polarizations can then be summarized,
\begin{equation}\label{eq:thinsheet}
(T, A, R)=\frac{(1, \myeta, \myeta^2/4)}{(1+\myeta/2)^2},
\end{equation}
where we have defined effective conductivity ratios
\begin{subequations}\label{eq:etas}
\begin{align}
\myeta_S&\equiv \frac{\sigma_{\text{2D}} Z_0}{n \cos\theta},\quad \text{and}\label{eq:etaS}\\
\myeta_P&\equiv \sigma_{\text{2D}} Z_0 \cos \theta/n\label{eq:etaP}
\end{align}
\end{subequations}
for $S$-polarization and  $P$-polarization respectively. Equations (\ref{eq:thinsheet}) and (\ref{eq:etas}), plotted in  Fig.~\ref{fig:thinsheet}, capture the main results of this paper.

To elucidate the implications of Eqs.~(\ref{eq:thinsheet}) and (\ref{eq:etas}) we take $n=1$ and first consider the case of normal incidence ($\cos\theta=1$). Even for the usual case of positive conductivity ($\myeta>0$), Eq.~(\ref{eq:thinsheet}) shows interesting behavior.  As expected, for vanishing conductivity the transmission $T\rightarrow 1$ and the reflection $R\rightarrow 0$, whereas for high conductivity $T\rightarrow 0$ and $R\rightarrow 1$.  The absorption $A$ vanishes in both limits, but the intermediate maximum is surprisingly large.  As has been noted previously  \cite{1934Woltersdorff,1964Kaplan,2003Bosman,2006Kaplan}, a sheet with half the impedance of free space absorbs 50\% of the incident power at normal incidence, even if it is thin compared to a skin depth.  The 50\% limit on the absorption results from the lack of dissipative coupling to the magnetic half of the electromagnetic wave's energy \cite{2014Yu}, but excepting this factor of two the coupling is maximal.

Proceeding to the case of non-normal incidence, we see that, because of the cosines in Eqs.~(\ref{eq:etas}), varying the polarization and angle of incidence  can accomplish the same result as varying the conductivity $\sigma_\text{2D}$.  
Working in $S$-polarization ($P$-polarization) away from normal incidence effectively increases (decreases) the sheet conductivity.
To the extent that glancing incidence ($\theta\rightarrow \pi/2$) can be achieved, any effective conductivity ratio $\myeta$ is possible regardless of the actual sheet conductivity $\sigma_\text{2D}$.  As a consequence, maximal coupling can be arranged in any one of the transmission, absorption, or reflection channels.

Particularly important is the geometry that gives strong coupling in the absorption channel. A sheet with, for example, low conductivity can be made to achieve maximal $A$ by arranging an angle of incidence $\theta=\arccos( \sigma_\text{2D} Z_0/2)$ in $S$-polarization.  If this impedance-matching condition is satisfied with positive conductivity, then $\myeta=2$ and $A=0.5$.   Such large absorption in a thin layer could help improve the economics of solar cells, for instance.

With an inverted, or negative-conductivity, sheet, achieving the corresponding maximal coupling condition, $\myeta=-2$, gives extraordinary behavior.  In this case $T$, $A$, and $R$ are all divergent, with $T$ and $R$ positive and $A$ negative (see Fig.~\ref{fig:thinsheet}).  As $\myeta\rightarrow -2$ the sheet effectively becomes an amplifying 50:50 beamsplitter with large gain in both the transmitted and the reflected beams.  

As in the case of positive conductivity, the effective conductivity ratio $\myeta$ can be tuned geometrically, perhaps to achieve a certain impedance match \cite{2014Yu}.  If for technical reasons creating a highly-conductive, inverted medium is difficult, larger negative $\myeta$ can still be achieved by working at glancing incidence in $S$-polarization. For both conductivity cases the geometric tuning is nontrivial, i.e., not a Lambertian cosine dependence resulting from changes in the sheet's projected area \cite{2010Tree}.

Normally one thinks of gain as being equivalent to negative absorption. While true, this viewpoint easily leads to the erroneous conclusion that weakly absorbing materials have limited potential for producing large, single-`pass' gain.  Here we might expect that the 50\% absorption limit would imply a maximum gain of $\times 1.5$ or $\times 2$. In contrast with this na\"ive expectation, the amplification produced by an inverted medium can be much larger than the corresponding attenuation produced by the uninverted medium.  Changing the sign of the conductivity is not equivalent to changing the sign of the absorption. 

Qualitatively the two-dimensional (2D), negative-conductivity sheet with $\myeta <-1$ has a unique feature:  it gives gain which cannot be accredited to the process of transmission through an inverted medium, but rather occurs during reflection at an interface.  A quantum-mechanical description of this process must necessarily identify reflection as a form of stimulated emission.  Typically stimulated emission is considered to create a new photon with quantum numbers identical to those of the incident photon, but with reflection this identity can no longer hold; one component of the new photon's momentum must have sign opposite that of the incident photon. 

The potential divergence seen in Eq.~(\ref{eq:thinsheet}) is reminiscent of the one occurring in a Fabry-Perot resonator with cavity gain greater than its round-trip losses, where it is indicative of self-oscillation \cite{1995Verdeyen}. In this analogy the amplifying beamsplitter corresponds to the zero-mode of a Fabry-Perot etalon, with no cavity or resonant structure. As in a laser, the pump process maintaining the negative conductivity only supplies energy at a limited rate, so in an actual physical implementation divergent gain will not be achieved.  The magnitude of $\myeta$ will decrease as the amplification rate approaches the pump rate. Nonetheless exploring the origin of this infinity is instructive. As is so often the case, the formal divergence here can be traced to an unphysical geometric assumption: we have taken the conducting sheet to have zero thickness, while requiring that its conductivity remain finite.  While at times a conducting sheet can be considered to be infinitely thin in the first approximation, this simplification gives quantitatively inaccurate results near $\myeta=-2$.

\section{Slab}

To treat the problem more realistically we apply the method of transfer matrices \cite{1999Jackson} to a slab of conducting material of finite thickness.  (We use the terms \emph{sheet} and \emph{slab} as  shorthand for the infinitesimally thin layer calculated above and the finite thickness case treated below respectively.)  Using this method the relationship between the incident, reflected, and transmitted electric fields can be written as  
\begin{equation}\label{eq:defineM}
\begin{pmatrix} E_t\\0\end{pmatrix}=M\begin{pmatrix}E_i\\E_r\end{pmatrix}.
\end{equation}
The $2\times2$ matrix $M$ effecting the transformation is built up from a series of matrices, each describing the transfer across an interface or through some thickness of a medium.  For the case of an incident wave from a medium (1) entering a second (2) and continuing on to a third (3), the transfer matrix has the form $M=\mathcal{I}_{23}\mathcal{T}_{2}\mathcal{I}_{12}$, where the translation matrix $\mathcal{T}$ and the interface matrices $\mathcal{I}$ account for the slab and boundaries designated by the respective subscripts. The procedure for generalizing such an expression to allow for an arbitrary number of finite-thickness slabs with different electromagnetic properties follows from induction and can be deduced by inspection.  The matrix that evolves the wave across the $j$th slab of thickness $d_j$ and index $n_j$ is
\begin{equation}\label{eq:TranslationMatrix}
\mathcal{T}_{j}=\begin{pmatrix}
e^{i\delta_j} & 0\\
0  & e^{-i\delta_j}
\end{pmatrix},
\end{equation}
where $\delta_j=(\omega/c) d_j n_j \cos \theta_j$. For $S$-polarization the matrix enforcing the boundary conditions at an interface between two distinct media $j$ and $k$ is
\begin{equation}\label{eq:InterfaceMatrixS}
\mathcal{I}^S_{jk}=\frac{1}{2}\begin{pmatrix}
1+\frac{n_j \cos\theta_j}{n_k \cos\theta_k} &  1-\frac{n_j \cos\theta_j}{n_k \cos\theta_k} \\
1-\frac{n_j \cos\theta_j}{n_k \cos\theta_k}  & 1+\frac{n_j \cos\theta_j}{n_k \cos\theta_k}
\end{pmatrix}.
\end{equation}
Here the $n$'s and the $\theta$'s refer to the refractive indices and propagation angles (determined by Snell's law) in the respective media. For $P$-polarization $\mathcal{I}^P_{jk}$ has a similar form, 
\begin{equation}\label{eq:InterfaceMatrixP}
\mathcal{I}^P_{jk}=\frac{1}{2}\begin{pmatrix}
\frac{n_j}{n_k}+\frac{\cos\theta_j}{\cos\theta_k} &  \frac{n_j}{n_k}-\frac{\cos\theta_j}{\cos\theta_k} \\
\frac{n_j}{n_k}-\frac{\cos\theta_j}{\cos\theta_k}  & \frac{n_j}{n_k}+\frac{\cos\theta_j}{\cos\theta_k}
\end{pmatrix}.
\end{equation}
Explicit though lengthy expressions for $T$, $A$, and $R$  are easily found using Eqs.~(\ref{eq:defineM})--(\ref{eq:InterfaceMatrixP}) given above.

To see the approach to  the case of a conducting sheet in vacuum discussed previously, we take $n_1=n_3$, and the slab's refractive index $n_2=\tilde{n}+i\tilde{k}$ to be given by
\begin{equation}\label{eq:ComplexPermittivity}
n^2_2=\hat{\varepsilon}_2\equiv \varepsilon_r+ i \varepsilon_i=\varepsilon_r+ i \frac{4 \pi \sigma_\text{3D}}{\omega}=\varepsilon_r+ i\frac{ \lambda}{2\pi} \sigma_\text{3D}Z_0,
\end{equation}
where $\hat{\varepsilon}_2$ is the complex permittivity and $\lambda$ is the vacuum wavelength of the incident radiation.  The three-dimensional (3D) conductivity $ \sigma_\text{3D} \rightarrow  \sigma_\text{2D}/d$ in the limit $d \rightarrow 0$.  For $S$-polarization, which is the more interesting case since invertible materials generally have small conductivities, 
\begin{equation}
(t_S,r_S)=\frac{(2x, i (x^2-1)\sin \delta_2)}{2 x \cos \delta_2 - i (x^2+1)\sin \delta_2}
\end{equation}
with $x\equiv n_2 \cos\theta_2/n_1 \cos\theta_1$. Assuming an index match ($\varepsilon_r=n_1^2$) and a small phase $\delta_2$ (i.e., a thin slab) gives
\begin{equation}
(t_S,r_S)\simeq\frac{(2,-\myeta')}{(2+\myeta')(1-a^2/2)-i a (2+\myeta'+\myeta'^2/6-a^2/3)},
\end{equation}
where $\myeta'\equiv\sigma_{\text{3D}} d_2 Z_0/n_1 \cos\theta_1$, $a\equiv(\omega/c) d_2 n_1 \cos\theta_1$, and we have kept terms to third order in $\delta_2$ in the denominator.  At $\myeta'=-2$, which corresponds to the infinite gain condition ($\myeta=-2$) of the sheet case, this expression gives, to leading order, $T=R\equiv G$, where 
\begin{equation}\label{eq:slabmax}
G\simeq\Big(\frac{3}{a}\Big)^2=\Big(\frac{3 \lambda}{ \pi d_2^2 \sigma_\text{3D} Z_0}\Big)^2=\Big(\frac{3 \varepsilon_i}{2 n_1^2 \cos^2\theta_1}\Big)^2.
\end{equation}
Arranging $\varepsilon_r-n_1^2\sin^2\theta_1= (1/3)n_1^2\cos^2\theta_1$ gives an even larger $G\simeq 25 (3/a)^4$, a result found by keeping terms to fifth order in $\delta_2$. Thus the gain $G$ from a slab is finite for non-vanishing thickness $d_2$, and diverges as $d_2\rightarrow 0$ with $d_2 \sigma_\text{3D}\rightarrow  \sigma_\text{2D}$ fixed.

The reflection gain of the slab arrangement (\ref{eq:slabmax}) contrasts with the analogous result for the interface of two semi-infinite media, where the second one is active.  There   $R\leq 1$ in all cases \cite{2010Siegman,2012Perez}.  An infinitely thick inverted medium produces zero gain in reflection, while a thin, inverted medium can produce large $R$.  

That a thin slab can produce large $T$ is also surprising, since in typical optical amplifiers the small signal gain (in transmission, of course)  grows exponentially as the gain medium thickness increases \cite{1995Verdeyen,2003Silfvast}. This last point highlights the utility of an extended concept of impedance matching for negative-conductivity systems.  Here, as a function of the conductivity and the thickness, the gain peaks at the negative analog of the best impedance match.%shrinks decreases

The role of impedance matching provides another perspective on the distinction between the amplifying beamsplitter and previously described optical amplifiers. The active mirror, for instance, is inherently a multipass transmission device \cite{1981Abate} that relies on an impedance mismatch to produce reflection. Likewise, amplifier designs based on total internal reflection, from fiber lasers \cite{1966Koester} to whispering-gallery-mode microsphere lasers \cite{1996Sandoghdar}, require an impedance mismatch in the form of an index discrepancy to produce reflection.  The amplifying beamsplitter works best in the limit where an impedance half match, modulo a sign, is achieved.

\add{\section{Real materials}}

Although arranging the inversion and geometry to give $\myeta=-2$ in a thin layer might be technically challenging, in a physical implementation with real materials the gain described by Eq.~(\ref{eq:slabmax}) could be made large.  As a first example we consider graphene, the canonical example of a thin conductor  or ``quantum membrane'' \cite{2013Fang}.  Graphene has a two-dimensional optical conductivity 
\begin{equation}\label{eq:tree}
\sigma_{\text{2D}}=(\pi \alpha/Z_0)(n_v-n_c),
\end{equation} 
where $\alpha \simeq 1/137$ is the fine structure constant, and $n_v$ and $n_c$ are the occupations of the valence and conduction bands respectively  \cite{2010Tree}.  If doping and thermal excitations are negligible, then $n_v=1$ and $n_c=0$ and this expression reduces to the famous  result connecting the optical conductivity $e^2/4\hbar$ to the absorption $\pi \alpha$ \cite{2002Ando,2008Nair}.  Partial inversion of graphene has been achieved with strong photoexcitation \cite{2012LiFemtosecond,2013Gierz}.  Assuming a complete inversion, a graphene thickness of 0.34~nm, and incident radiation with $\lambda=800$~nm, Eq.~(\ref{eq:slabmax}) gives gains in transmission and reflection of $\sim 10^{10}$. \add{While such an inversion is unlikely to be realized in graphene, transition-metal dichalcogenides and other 2D layered semiconductors (e.g., WSe$_2$) are similarly thin, with similar conductivities} \cite{2013Fang}\add{, and can support practical inversions} \cite{2015Wu}. \add{Thus such direct band gap materials show promise for realizing large gain.}

Compared to many invertible materials \change{graphene is a good conductor, but it}{these 2D layered semiconductors are good conductors at optical frequencies, but graphene --- to continue using this example ---} only achieves the maximum coupling condition in $S$-polarization at the glancing angle $\theta_1\simeq 89.3^\circ$.  If low conductivity makes $\theta_1$ inconveniently close to $90^\circ$, the angle of incidence corresponding to maximal coupling can be decreased by working in very low index materials, e.g., photonic crystals \cite{1994Dowling,2000Gralak}, or by increasing the ratio of $n_1$ to $n_3$.  Moving into the total internal reflection regime, however, gives an impedance mismatch that destroys the beamsplitter and spoils the system gain. 

Standard invertible materials are characterized by their gain coefficients $g$, which are related to optical conductivities by  $g\simeq \sigma_\text{3D} Z_0/n$ when $\varepsilon_i\ll \varepsilon_r$ [see Eq.~\ref{eq:ComplexPermittivity}] \cite{1995Heald}.  Ruby, neodymium-doped yttrium aluminum garnet (Nd:YAG), neodymium-doped yttrium orthovanadate (Nd:YVO$_4$), and titanium-doped sapphire have gain coefficients in the range 1--10 cm$^{-1}$  \cite{2003Silfvast,1994Bernard}, making them more than $10^4$ times less conductive than graphene.  The resulting angles of incidence are so near $90^\circ$ that a physical optics picture may be required.  However, semiconducting materials can have gain coefficients as large as $10^4$--$10^5$ cm$^{-1}$  \cite{1973Shaklee}\add{, and crystalline monolayers are starting to see use as laser gain media} \cite{2015Wu}.  Taking 20~nm of index-matched, low-temperature gallium nitride (GaN, $n=2.5$, $g=10^5$ cm$^{-1}$, $\lambda=359$~nm) as an example \cite{1971Dingle}, we find $G\simeq 1200$ at $\theta\simeq 84.3^\circ$, an angle that can be straightforwardly arranged.  Decreasing the thickness or the gain coefficient gives larger gain at a larger angle; $d=5$~nm gives $G\simeq 2.9\times 10^5$ at $\theta\simeq 88.6^\circ$, while $g=10^4$ cm$^{-1}$ gives $G\simeq 1.2\times 10^5$ at $\theta\simeq 89.4^\circ$. Known materials have the optical properties required to create a thin ($\delta_2 \ll 1$), amplifying beamsplitter with large gain. 

Despite appearances in these glancing  incidence examples, waveguide effects are not involved: \add{with} a good \add{inverted} conductor \add{one} could achieve large gain at normal incidence.  Lead telluride (PbTe) has $\hat{\varepsilon}\simeq 0.33+ 50 i$ near $\lambda=620$~nm \cite{1994Suzuki}, corresponding to a skin depth $\lambda/2\pi \tilde{k}\simeq 20$~nm. \change{Such}{At normal incidence such} material in a slab of thickness $d=\lambda/\pi \varepsilon_i\simeq 3.9$~nm in vacuum would give \change{50\% absorption normally}{$A=50$\%}, and $G>10^8$ \add{if it could be} fully inverted.  As mentioned earlier, saturation effects limit the gain in practical situations, so such large $G$ should be taken to indicate nearly ideal coupling only.

\begin{acknowledgments}
This work has been supported by NSF award DMR-1206849.\vspace{-0.1 in}
\end{acknowledgments}

%Comment the next 4 lines for non-review version
%\vspace{0.2 in}
%\textbf{References are given with titles in this for-review copy only.}
%\vspace{-0.1 in}
%\bibliographystyle{naturemag19}
\bibliography{blackgrapheneZ}

\begin{thebibliography}{29}%
\makeatletter
\providecommand \@ifxundefined [1]{%
 \@ifx{#1\undefined}
}%
\providecommand \@ifnum [1]{%
 \ifnum #1\expandafter \@firstoftwo
 \else \expandafter \@secondoftwo
 \fi
}%
\providecommand \@ifx [1]{%
 \ifx #1\expandafter \@firstoftwo
 \else \expandafter \@secondoftwo
 \fi
}%
\providecommand \natexlab [1]{#1}%
\providecommand \enquote  [1]{``#1''}%
\providecommand \bibnamefont  [1]{#1}%
\providecommand \bibfnamefont [1]{#1}%
\providecommand \citenamefont [1]{#1}%
\providecommand \href@noop [0]{\@secondoftwo}%
\providecommand \href [0]{\begingroup \@sanitize@url \@href}%
\providecommand \@href[1]{\@@startlink{#1}\@@href}%
\providecommand \@@href[1]{\endgroup#1\@@endlink}%
\providecommand \@sanitize@url [0]{\catcode `\\12\catcode `\$12\catcode
  `\&12\catcode `\#12\catcode `\^12\catcode `\_12\catcode `\%12\relax}%
\providecommand \@@startlink[1]{}%
\providecommand \@@endlink[0]{}%
\providecommand \url  [0]{\begingroup\@sanitize@url \@url }%
\providecommand \@url [1]{\endgroup\@href {#1}{\urlprefix }}%
\providecommand \urlprefix  [0]{URL }%
\providecommand \Eprint [0]{\href }%
\providecommand \doibase [0]{http://dx.doi.org/}%
\providecommand \selectlanguage [0]{\@gobble}%
\providecommand \bibinfo  [0]{\@secondoftwo}%
\providecommand \bibfield  [0]{\@secondoftwo}%
\providecommand \translation [1]{[#1]}%
\providecommand \BibitemOpen [0]{}%
\providecommand \bibitemStop [0]{}%
\providecommand \bibitemNoStop [0]{.\EOS\space}%
\providecommand \EOS [0]{\spacefactor3000\relax}%
\providecommand \BibitemShut  [1]{\csname bibitem#1\endcsname}%
\let\auto@bib@innerbib\@empty
%</preamble>
\bibitem [{\citenamefont {Heald}\ and\ \citenamefont
  {Marion}(1995)}]{1995Heald}%
  \BibitemOpen
  \bibfield  {author} {\bibinfo {author} {\bibfnamefont {Mark~A.}\ \bibnamefont
  {Heald}}\ and\ \bibinfo {author} {\bibfnamefont {Jerry~B.}\ \bibnamefont
  {Marion}},\ }\href@noop {} {\emph {\bibinfo {title} {Classical
  Electromagnetic Radiation}}}\ (\bibinfo  {publisher} {{Saunders College
  Pub.}},\ \bibinfo {address} {Fort Worth},\ \bibinfo {year}
  {1995})\BibitemShut {NoStop}%
\bibitem [{\citenamefont {Veselago}(1968)}]{1968Veselago}%
  \BibitemOpen
  \bibfield  {author} {\bibinfo {author} {\bibfnamefont {Viktor~G}\
  \bibnamefont {Veselago}},\ }\bibfield  {title} {\enquote {\bibinfo {title}
  {The electrodynamics of substances with simultaneously negative values of
  $\epsilon$ and $\mu$},}\ }\href@noop {} {\bibfield  {journal} {\bibinfo
  {journal} {Soviet Physics Uspekhi}\ }\textbf {\bibinfo {volume} {10}},\
  \bibinfo {pages} {509} (\bibinfo {year} {1968})}\BibitemShut {NoStop}%
\bibitem [{\citenamefont {Pendry}(2000)}]{2000Pendry}%
  \BibitemOpen
  \bibfield  {author} {\bibinfo {author} {\bibfnamefont {J.~B.}\ \bibnamefont
  {Pendry}},\ }\bibfield  {title} {\enquote {\bibinfo {title} {Negative
  {{Refraction Makes}} a {{Perfect Lens}}},}\ }\href@noop {} {\bibfield
  {journal} {\bibinfo  {journal} {Physical Review Letters}\ }\textbf {\bibinfo
  {volume} {85}},\ \bibinfo {pages} {3966--3969} (\bibinfo {year}
  {2000})}\BibitemShut {NoStop}%
\bibitem [{\citenamefont {Abate}\ \emph {et~al.}(1981)\citenamefont {Abate},
  \citenamefont {Lund}, \citenamefont {Brown}, \citenamefont {Jacobs},
  \citenamefont {Refermat}, \citenamefont {Kelly}, \citenamefont {Gavin},
  \citenamefont {Waldbillig},\ and\ \citenamefont {Lewis}}]{1981Abate}%
  \BibitemOpen
  \bibfield  {author} {\bibinfo {author} {\bibfnamefont {J.~A.}\ \bibnamefont
  {Abate}}, \bibinfo {author} {\bibfnamefont {L.}~\bibnamefont {Lund}},
  \bibinfo {author} {\bibfnamefont {D.}~\bibnamefont {Brown}}, \bibinfo
  {author} {\bibfnamefont {S.}~\bibnamefont {Jacobs}}, \bibinfo {author}
  {\bibfnamefont {S.}~\bibnamefont {Refermat}}, \bibinfo {author}
  {\bibfnamefont {J.}~\bibnamefont {Kelly}}, \bibinfo {author} {\bibfnamefont
  {M.}~\bibnamefont {Gavin}}, \bibinfo {author} {\bibfnamefont
  {J.}~\bibnamefont {Waldbillig}}, \ and\ \bibinfo {author} {\bibfnamefont
  {O.}~\bibnamefont {Lewis}},\ }\bibfield  {title} {\enquote {\bibinfo {title}
  {Active mirror: A large-aperture medium-repetition rate {{Nd}}:glass
  amplifier},}\ }\href@noop {} {\bibfield  {journal} {\bibinfo  {journal}
  {Applied Optics}\ }\textbf {\bibinfo {volume} {20}},\ \bibinfo {pages}
  {351--361} (\bibinfo {year} {1981})}\BibitemShut {NoStop}%
\bibitem [{\citenamefont {Koester}(1966)}]{1966Koester}%
  \BibitemOpen
  \bibfield  {author} {\bibinfo {author} {\bibfnamefont {C.}~\bibnamefont
  {Koester}},\ }\bibfield  {title} {\enquote {\bibinfo {title} {{{9A4}} -
  {{Laser}} action by enhanced total internal reflection},}\ }\href@noop {}
  {\bibfield  {journal} {\bibinfo  {journal} {IEEE Journal of Quantum
  Electronics}\ }\textbf {\bibinfo {volume} {2}},\ \bibinfo {pages} {580--584}
  (\bibinfo {year} {1966})}\BibitemShut {NoStop}%
\bibitem [{\citenamefont {Woltersdorff}(1934)}]{1934Woltersdorff}%
  \BibitemOpen
  \bibfield  {author} {\bibinfo {author} {\bibfnamefont {Wilhelm}\ \bibnamefont
  {Woltersdorff}},\ }\bibfield  {title} {\enquote {\bibinfo {title} {{{\"U}ber
  die optischen Konstanten d{\"u}nner Metallschichten im langwelligen
  Ultrarot}},}\ }\href {\doibase 10.1007/BF01341647} {\bibfield  {journal}
  {\bibinfo  {journal} {Zeitschrift f{\"u}r Physik}\ }\textbf {\bibinfo
  {volume} {91}},\ \bibinfo {pages} {230--252} (\bibinfo {year}
  {1934})}\BibitemShut {NoStop}%
\bibitem [{\citenamefont {Kaplan}(1964)}]{1964Kaplan}%
  \BibitemOpen
  \bibfield  {author} {\bibinfo {author} {\bibfnamefont {A.~E.}\ \bibnamefont
  {Kaplan}},\ }\bibfield  {title} {\enquote {\bibinfo {title} {On the
  reflectivity of metallic films at microwave and radio frequencies},}\
  }\href@noop {} {\bibfield  {journal} {\bibinfo  {journal} {Radio Engineering
  and Electronic Physics}\ }\textbf {\bibinfo {volume} {9}},\ \bibinfo {pages}
  {1476--1481} (\bibinfo {year} {1964})}\BibitemShut {NoStop}%
\bibitem [{\citenamefont {Bosman}\ \emph {et~al.}(2003)\citenamefont {Bosman},
  \citenamefont {Lau},\ and\ \citenamefont {Gilgenbach}}]{2003Bosman}%
  \BibitemOpen
  \bibfield  {author} {\bibinfo {author} {\bibfnamefont {H.}~\bibnamefont
  {Bosman}}, \bibinfo {author} {\bibfnamefont {Y.~Y.}\ \bibnamefont {Lau}}, \
  and\ \bibinfo {author} {\bibfnamefont {R.~M.}\ \bibnamefont {Gilgenbach}},\
  }\bibfield  {title} {\enquote {\bibinfo {title} {Microwave absorption on a
  thin film},}\ }\href {\doibase 10.1063/1.1556969} {\bibfield  {journal}
  {\bibinfo  {journal} {Applied Physics Letters}\ }\textbf {\bibinfo {volume}
  {82}},\ \bibinfo {pages} {1353--1355} (\bibinfo {year} {2003})}\BibitemShut
  {NoStop}%
\bibitem [{\citenamefont {Kaplan}\ and\ \citenamefont
  {Zeldovich}(2006)}]{2006Kaplan}%
  \BibitemOpen
  \bibfield  {author} {\bibinfo {author} {\bibfnamefont {A.~E.}\ \bibnamefont
  {Kaplan}}\ and\ \bibinfo {author} {\bibfnamefont {B.~Ya.}\ \bibnamefont
  {Zeldovich}},\ }\bibfield  {title} {\enquote {\bibinfo {title} {Free-space
  terminator and coherent broadband blackbody interferometry},}\ }\href
  {\doibase 10.1364/OL.31.000335} {\bibfield  {journal} {\bibinfo  {journal}
  {Optics Letters}\ }\textbf {\bibinfo {volume} {31}},\ \bibinfo {pages} {335}
  (\bibinfo {year} {2006})}\BibitemShut {NoStop}%
\bibitem [{\citenamefont {Yu}\ and\ \citenamefont {Capasso}(2014)}]{2014Yu}%
  \BibitemOpen
  \bibfield  {author} {\bibinfo {author} {\bibfnamefont {Nanfang}\ \bibnamefont
  {Yu}}\ and\ \bibinfo {author} {\bibfnamefont {Federico}\ \bibnamefont
  {Capasso}},\ }\bibfield  {title} {\enquote {\bibinfo {title} {Flat optics
  with designer metasurfaces},}\ }\href {\doibase 10.1038/nmat3839} {\bibfield
  {journal} {\bibinfo  {journal} {Nature Materials}\ }\textbf {\bibinfo
  {volume} {13}},\ \bibinfo {pages} {139--150} (\bibinfo {year}
  {2014})}\BibitemShut {NoStop}%
\bibitem [{\citenamefont {Mecklenburg}\ \emph {et~al.}(2010)\citenamefont
  {Mecklenburg}, \citenamefont {Woo},\ and\ \citenamefont {Regan}}]{2010Tree}%
  \BibitemOpen
  \bibfield  {author} {\bibinfo {author} {\bibfnamefont {Matthew}\ \bibnamefont
  {Mecklenburg}}, \bibinfo {author} {\bibfnamefont {Jason}\ \bibnamefont
  {Woo}}, \ and\ \bibinfo {author} {\bibfnamefont {B.~C.}\ \bibnamefont
  {Regan}},\ }\bibfield  {title} {\enquote {\bibinfo {title} {Tree-level
  electron-photon interactions in graphene},}\ }\href {\doibase
  10.1103/PhysRevB.81.245401} {\bibfield  {journal} {\bibinfo  {journal}
  {Physical Review B}\ }\textbf {\bibinfo {volume} {81}},\ \bibinfo {pages}
  {245401} (\bibinfo {year} {2010})}\BibitemShut {NoStop}%
\bibitem [{\citenamefont {Verdeyen}(1995)}]{1995Verdeyen}%
  \BibitemOpen
  \bibfield  {author} {\bibinfo {author} {\bibfnamefont {Joseph~T.}\
  \bibnamefont {Verdeyen}},\ }\href@noop {} {\emph {\bibinfo {title} {Laser
  {{Electronics}}}}},\ \bibinfo {edition} {3rd}\ ed.\ (\bibinfo  {publisher}
  {{Prentice Hall}},\ \bibinfo {address} {Englewood Cliffs, N.J},\ \bibinfo
  {year} {1995})\BibitemShut {NoStop}%
\bibitem [{\citenamefont {Jackson}(1999)}]{1999Jackson}%
  \BibitemOpen
  \bibfield  {author} {\bibinfo {author} {\bibfnamefont {John~David}\
  \bibnamefont {Jackson}},\ }\href@noop {} {\emph {\bibinfo {title} {Classical
  Electrodynamics}}},\ \bibinfo {edition} {3rd}\ ed.\ (\bibinfo  {publisher}
  {{Wiley}},\ \bibinfo {address} {New York},\ \bibinfo {year}
  {1999})\BibitemShut {NoStop}%
\bibitem [{\citenamefont {Siegman}(2010)}]{2010Siegman}%
  \BibitemOpen
  \bibfield  {author} {\bibinfo {author} {\bibfnamefont {Anthony}\ \bibnamefont
  {Siegman}},\ }\bibfield  {title} {\enquote {\bibinfo {title} {Fresnel
  {{Reflection}}, {{Lenserf Reflection}} and {{Evanescent Gain}}},}\ }\href
  {\doibase 10.1364/OPN.21.1.000038} {\bibfield  {journal} {\bibinfo  {journal}
  {Optics and Photonics News}\ }\textbf {\bibinfo {volume} {21}},\ \bibinfo
  {pages} {38--45} (\bibinfo {year} {2010})}\BibitemShut {NoStop}%
\bibitem [{\citenamefont {Perez}\ \emph {et~al.}(2012)\citenamefont {Perez},
  \citenamefont {Matteo}, \citenamefont {Etcheverry},\ and\ \citenamefont
  {Dupla{\'a}}}]{2012Perez}%
  \BibitemOpen
  \bibfield  {author} {\bibinfo {author} {\bibfnamefont {Liliana~I.}\
  \bibnamefont {Perez}}, \bibinfo {author} {\bibfnamefont {Claudia~L.}\
  \bibnamefont {Matteo}}, \bibinfo {author} {\bibfnamefont {Javier}\
  \bibnamefont {Etcheverry}}, \ and\ \bibinfo {author} {\bibfnamefont
  {Mar{\'\i}a~Celeste}\ \bibnamefont {Dupla{\'a}}},\ }\bibfield  {title}
  {\enquote {\bibinfo {title} {Active isotropic slabs: Conditions for amplified
  reflection},}\ }\href {\doibase 10.1088/2040-8978/14/12/125711} {\bibfield
  {journal} {\bibinfo  {journal} {Journal of Optics}\ }\textbf {\bibinfo
  {volume} {14}},\ \bibinfo {pages} {125711} (\bibinfo {year}
  {2012})}\BibitemShut {NoStop}%
\bibitem [{\citenamefont {Silfvast}(2003)}]{2003Silfvast}%
  \BibitemOpen
  \bibfield  {author} {\bibinfo {author} {\bibfnamefont {William~T.}\
  \bibnamefont {Silfvast}},\ }\bibfield  {title} {\enquote {\bibinfo {title}
  {Lasers},}\ }in\ \href@noop {} {\emph {\bibinfo {booktitle} {Fundamentals of
  {{Photonics}}}}},\ \bibinfo {editor} {edited by\ \bibinfo {editor}
  {\bibfnamefont {Arthur}\ \bibnamefont {Guenther}}, \bibinfo {editor}
  {\bibfnamefont {Leno~S.}\ \bibnamefont {Pedrotti}}, \ and\ \bibinfo {editor}
  {\bibfnamefont {Chandrasekhar}\ \bibnamefont {Roychoudhuri}}}\ (\bibinfo
  {publisher} {{SPIE}},\ \bibinfo {address} {Bellingham},\ \bibinfo {year}
  {2003})\ pp.\ \bibinfo {pages} {1--45}\BibitemShut {NoStop}%
\bibitem [{\citenamefont {Sandoghdar}\ \emph {et~al.}(1996)\citenamefont
  {Sandoghdar}, \citenamefont {Treussart}, \citenamefont {Hare}, \citenamefont
  {Lefevre-Seguin}, \citenamefont {Raimond},\ and\ \citenamefont
  {Haroche}}]{1996Sandoghdar}%
  \BibitemOpen
  \bibfield  {author} {\bibinfo {author} {\bibfnamefont {V.}~\bibnamefont
  {Sandoghdar}}, \bibinfo {author} {\bibfnamefont {F.}~\bibnamefont
  {Treussart}}, \bibinfo {author} {\bibfnamefont {J.}~\bibnamefont {Hare}},
  \bibinfo {author} {\bibfnamefont {V.}~\bibnamefont {Lefevre-Seguin}},
  \bibinfo {author} {\bibfnamefont {J.~M.}\ \bibnamefont {Raimond}}, \ and\
  \bibinfo {author} {\bibfnamefont {S.}~\bibnamefont {Haroche}},\ }\bibfield
  {title} {\enquote {\bibinfo {title} {Very low threshold
  whispering-gallery-mode microsphere laser},}\ }\href@noop {} {\bibfield
  {journal} {\bibinfo  {journal} {Physical Review A}\ }\textbf {\bibinfo
  {volume} {54}},\ \bibinfo {pages} {R1777--R1780} (\bibinfo {year}
  {1996})}\BibitemShut {NoStop}%
\bibitem [{\citenamefont {Fang}\ \emph {et~al.}(2013)\citenamefont {Fang},
  \citenamefont {Bechtel}, \citenamefont {Plis}, \citenamefont {Martin},
  \citenamefont {Krishna}, \citenamefont {Yablonovitch},\ and\ \citenamefont
  {Javey}}]{2013Fang}%
  \BibitemOpen
  \bibfield  {author} {\bibinfo {author} {\bibfnamefont {Hui}\ \bibnamefont
  {Fang}}, \bibinfo {author} {\bibfnamefont {Hans~A.}\ \bibnamefont {Bechtel}},
  \bibinfo {author} {\bibfnamefont {Elena}\ \bibnamefont {Plis}}, \bibinfo
  {author} {\bibfnamefont {Michael~C.}\ \bibnamefont {Martin}}, \bibinfo
  {author} {\bibfnamefont {Sanjay}\ \bibnamefont {Krishna}}, \bibinfo {author}
  {\bibfnamefont {Eli}\ \bibnamefont {Yablonovitch}}, \ and\ \bibinfo {author}
  {\bibfnamefont {Ali}\ \bibnamefont {Javey}},\ }\bibfield  {title} {\enquote
  {\bibinfo {title} {Quantum of optical absorption in two-dimensional
  semiconductors},}\ }\href {\doibase 10.1073/pnas.1309563110} {\bibfield
  {journal} {\bibinfo  {journal} {Proceedings of the National Academy of
  Sciences}\ }\textbf {\bibinfo {volume} {110}},\ \bibinfo {pages}
  {11688--11691} (\bibinfo {year} {2013})}\BibitemShut {NoStop}%
\bibitem [{\citenamefont {Ando}\ \emph {et~al.}(2002)\citenamefont {Ando},
  \citenamefont {Zheng},\ and\ \citenamefont {Suzuura}}]{2002Ando}%
  \BibitemOpen
  \bibfield  {author} {\bibinfo {author} {\bibfnamefont {T.}~\bibnamefont
  {Ando}}, \bibinfo {author} {\bibfnamefont {Y.~S.}\ \bibnamefont {Zheng}}, \
  and\ \bibinfo {author} {\bibfnamefont {H.}~\bibnamefont {Suzuura}},\
  }\bibfield  {title} {\enquote {\bibinfo {title} {Dynamical conductivity and
  zero-mode anomaly in honeycomb lattices},}\ }\href@noop {} {\bibfield
  {journal} {\bibinfo  {journal} {Journal of the Physical Society of Japan}\
  }\textbf {\bibinfo {volume} {71}},\ \bibinfo {pages} {1318--1324} (\bibinfo
  {year} {2002})}\BibitemShut {NoStop}%
\bibitem [{\citenamefont {Nair}\ \emph {et~al.}(2008)\citenamefont {Nair},
  \citenamefont {Blake}, \citenamefont {Grigorenko}, \citenamefont {Novoselov},
  \citenamefont {Booth}, \citenamefont {Stauber}, \citenamefont {Peres},\ and\
  \citenamefont {Geim}}]{2008Nair}%
  \BibitemOpen
  \bibfield  {author} {\bibinfo {author} {\bibfnamefont {R.~R.}\ \bibnamefont
  {Nair}}, \bibinfo {author} {\bibfnamefont {P.}~\bibnamefont {Blake}},
  \bibinfo {author} {\bibfnamefont {A.~N.}\ \bibnamefont {Grigorenko}},
  \bibinfo {author} {\bibfnamefont {K.~S.}\ \bibnamefont {Novoselov}}, \bibinfo
  {author} {\bibfnamefont {T.~J.}\ \bibnamefont {Booth}}, \bibinfo {author}
  {\bibfnamefont {T.}~\bibnamefont {Stauber}}, \bibinfo {author} {\bibfnamefont
  {N.~M.~R.}\ \bibnamefont {Peres}}, \ and\ \bibinfo {author} {\bibfnamefont
  {A.~K.}\ \bibnamefont {Geim}},\ }\bibfield  {title} {\enquote {\bibinfo
  {title} {Fine structure constant defines visual transparency of graphene},}\
  }\href@noop {} {\bibfield  {journal} {\bibinfo  {journal} {Science}\ }\textbf
  {\bibinfo {volume} {320}},\ \bibinfo {pages} {1308--1308} (\bibinfo {year}
  {2008})}\BibitemShut {NoStop}%
\bibitem [{\citenamefont {Li}\ \emph {et~al.}(2012)\citenamefont {Li},
  \citenamefont {Luo}, \citenamefont {Hupalo}, \citenamefont {Zhang},
  \citenamefont {Tringides}, \citenamefont {Schmalian},\ and\ \citenamefont
  {Wang}}]{2012LiFemtosecond}%
  \BibitemOpen
  \bibfield  {author} {\bibinfo {author} {\bibfnamefont {T.}~\bibnamefont
  {Li}}, \bibinfo {author} {\bibfnamefont {L.}~\bibnamefont {Luo}}, \bibinfo
  {author} {\bibfnamefont {M.}~\bibnamefont {Hupalo}}, \bibinfo {author}
  {\bibfnamefont {J.}~\bibnamefont {Zhang}}, \bibinfo {author} {\bibfnamefont
  {M.~C.}\ \bibnamefont {Tringides}}, \bibinfo {author} {\bibfnamefont
  {J.}~\bibnamefont {Schmalian}}, \ and\ \bibinfo {author} {\bibfnamefont
  {J.}~\bibnamefont {Wang}},\ }\bibfield  {title} {\enquote {\bibinfo {title}
  {Femtosecond {{Population Inversion}} and {{Stimulated Emission}} of {{Dense
  Dirac Fermions}} in {{Graphene}}},}\ }\href@noop {} {\bibfield  {journal}
  {\bibinfo  {journal} {Physical Review Letters}\ }\textbf {\bibinfo {volume}
  {108}},\ \bibinfo {pages} {167401} (\bibinfo {year} {2012})}\BibitemShut
  {NoStop}%
\bibitem [{\citenamefont {Gierz}\ \emph {et~al.}(2013)\citenamefont {Gierz},
  \citenamefont {Petersen}, \citenamefont {Mitrano}, \citenamefont {Cacho},
  \citenamefont {Turcu}, \citenamefont {Springate}, \citenamefont {St{\"o}hr},
  \citenamefont {K{\"o}hler}, \citenamefont {Starke},\ and\ \citenamefont
  {Cavalleri}}]{2013Gierz}%
  \BibitemOpen
  \bibfield  {author} {\bibinfo {author} {\bibfnamefont {Isabella}\
  \bibnamefont {Gierz}}, \bibinfo {author} {\bibfnamefont {Jesse~C.}\
  \bibnamefont {Petersen}}, \bibinfo {author} {\bibfnamefont {Matteo}\
  \bibnamefont {Mitrano}}, \bibinfo {author} {\bibfnamefont {Cephise}\
  \bibnamefont {Cacho}}, \bibinfo {author} {\bibfnamefont {I.~C.~Edmond}\
  \bibnamefont {Turcu}}, \bibinfo {author} {\bibfnamefont {Emma}\ \bibnamefont
  {Springate}}, \bibinfo {author} {\bibfnamefont {Alexander}\ \bibnamefont
  {St{\"o}hr}}, \bibinfo {author} {\bibfnamefont {Axel}\ \bibnamefont
  {K{\"o}hler}}, \bibinfo {author} {\bibfnamefont {Ulrich}\ \bibnamefont
  {Starke}}, \ and\ \bibinfo {author} {\bibfnamefont {Andrea}\ \bibnamefont
  {Cavalleri}},\ }\bibfield  {title} {\enquote {\bibinfo {title} {Snapshots of
  non-equilibrium {{Dirac}} carrier distributions in graphene},}\ }\href
  {\doibase 10.1038/nmat3757} {\bibfield  {journal} {\bibinfo  {journal}
  {Nature Materials}\ }\textbf {\bibinfo {volume} {12}},\ \bibinfo {pages}
  {1119--1124} (\bibinfo {year} {2013})}\BibitemShut {NoStop}%
\bibitem [{\citenamefont {Wu}\ \emph {et~al.}(2015)\citenamefont {Wu},
  \citenamefont {Buckley}, \citenamefont {Schaibley}, \citenamefont {Feng},
  \citenamefont {Yan}, \citenamefont {Mandrus}, \citenamefont {Hatami},
  \citenamefont {Yao}, \citenamefont {Vu{\v c}kovi{\'c}}, \citenamefont
  {Majumdar},\ and\ \citenamefont {Xu}}]{2015Wu}%
  \BibitemOpen
  \bibfield  {author} {\bibinfo {author} {\bibfnamefont {Sanfeng}\ \bibnamefont
  {Wu}}, \bibinfo {author} {\bibfnamefont {Sonia}\ \bibnamefont {Buckley}},
  \bibinfo {author} {\bibfnamefont {John~R.}\ \bibnamefont {Schaibley}},
  \bibinfo {author} {\bibfnamefont {Liefeng}\ \bibnamefont {Feng}}, \bibinfo
  {author} {\bibfnamefont {Jiaqiang}\ \bibnamefont {Yan}}, \bibinfo {author}
  {\bibfnamefont {David~G.}\ \bibnamefont {Mandrus}}, \bibinfo {author}
  {\bibfnamefont {Fariba}\ \bibnamefont {Hatami}}, \bibinfo {author}
  {\bibfnamefont {Wang}\ \bibnamefont {Yao}}, \bibinfo {author} {\bibfnamefont
  {Jelena}\ \bibnamefont {Vu{\v c}kovi{\'c}}}, \bibinfo {author} {\bibfnamefont
  {Arka}\ \bibnamefont {Majumdar}}, \ and\ \bibinfo {author} {\bibfnamefont
  {Xiaodong}\ \bibnamefont {Xu}},\ }\bibfield  {title} {\enquote {\bibinfo
  {title} {Monolayer semiconductor nanocavity lasers with ultralow
  thresholds},}\ }\href {\doibase 10.1038/nature14290} {\bibfield  {journal}
  {\bibinfo  {journal} {Nature}\ }\textbf {\bibinfo {volume} {520}},\ \bibinfo
  {pages} {69--72} (\bibinfo {year} {2015})}\BibitemShut {NoStop}%
\bibitem [{\citenamefont {Dowling}\ and\ \citenamefont
  {Bowden}(1994)}]{1994Dowling}%
  \BibitemOpen
  \bibfield  {author} {\bibinfo {author} {\bibfnamefont {Jonathan~P.}\
  \bibnamefont {Dowling}}\ and\ \bibinfo {author} {\bibfnamefont {Charles~M.}\
  \bibnamefont {Bowden}},\ }\bibfield  {title} {\enquote {\bibinfo {title}
  {Anomalous {{Index}} of {{Refraction}} in {{Photonic Bandgap Materials}}},}\
  }\href {\doibase 10.1080/09500349414550371} {\bibfield  {journal} {\bibinfo
  {journal} {Journal of Modern Optics}\ }\textbf {\bibinfo {volume} {41}},\
  \bibinfo {pages} {345--351} (\bibinfo {year} {1994})}\BibitemShut {NoStop}%
\bibitem [{\citenamefont {Gralak}\ \emph {et~al.}(2000)\citenamefont {Gralak},
  \citenamefont {Enoch},\ and\ \citenamefont {Tayeb}}]{2000Gralak}%
  \BibitemOpen
  \bibfield  {author} {\bibinfo {author} {\bibfnamefont {Boris}\ \bibnamefont
  {Gralak}}, \bibinfo {author} {\bibfnamefont {Stefan}\ \bibnamefont {Enoch}},
  \ and\ \bibinfo {author} {\bibfnamefont {G{\'e}rard}\ \bibnamefont {Tayeb}},\
  }\bibfield  {title} {\enquote {\bibinfo {title} {Anomalous refractive
  properties of photonic crystals},}\ }\href {\doibase 10.1364/JOSAA.17.001012}
  {\bibfield  {journal} {\bibinfo  {journal} {Journal of the Optical Society of
  America A}\ }\textbf {\bibinfo {volume} {17}},\ \bibinfo {pages} {1012--1020}
  (\bibinfo {year} {2000})}\BibitemShut {NoStop}%
\bibitem [{\citenamefont {Bernard}\ \emph {et~al.}(1994)\citenamefont
  {Bernard}, \citenamefont {McCullough},\ and\ \citenamefont
  {Alcock}}]{1994Bernard}%
  \BibitemOpen
  \bibfield  {author} {\bibinfo {author} {\bibfnamefont {J.~E.}\ \bibnamefont
  {Bernard}}, \bibinfo {author} {\bibfnamefont {E.}~\bibnamefont {McCullough}},
  \ and\ \bibinfo {author} {\bibfnamefont {A.~J.}\ \bibnamefont {Alcock}},\
  }\bibfield  {title} {\enquote {\bibinfo {title} {High gain, diode-pumped
  {{Nd}}:{{YVO$_4$}} slab amplifier},}\ }\href {\doibase
  10.1016/0030-4018(94)90746-3} {\bibfield  {journal} {\bibinfo  {journal}
  {Optics Communications}\ }\textbf {\bibinfo {volume} {109}},\ \bibinfo
  {pages} {109--114} (\bibinfo {year} {1994})}\BibitemShut {NoStop}%
\bibitem [{\citenamefont {Shaklee}\ \emph {et~al.}(1973)\citenamefont
  {Shaklee}, \citenamefont {Nahory},\ and\ \citenamefont
  {Leheny}}]{1973Shaklee}%
  \BibitemOpen
  \bibfield  {author} {\bibinfo {author} {\bibfnamefont {KL}~\bibnamefont
  {Shaklee}}, \bibinfo {author} {\bibfnamefont {RE}~\bibnamefont {Nahory}}, \
  and\ \bibinfo {author} {\bibfnamefont {RF}~\bibnamefont {Leheny}},\
  }\bibfield  {title} {\enquote {\bibinfo {title} {Optical gain in
  semiconductors},}\ }\href {\doibase 10.1016/0022-2313(73)90072-0} {\bibfield
  {journal} {\bibinfo  {journal} {Journal of Luminescence}\ }\textbf {\bibinfo
  {volume} {7}},\ \bibinfo {pages} {284--309} (\bibinfo {year}
  {1973})}\BibitemShut {NoStop}%
\bibitem [{\citenamefont {Dingle}\ \emph {et~al.}(1971)\citenamefont {Dingle},
  \citenamefont {Shaklee}, \citenamefont {Leheny},\ and\ \citenamefont
  {Zetterstrom}}]{1971Dingle}%
  \BibitemOpen
  \bibfield  {author} {\bibinfo {author} {\bibfnamefont {R}~\bibnamefont
  {Dingle}}, \bibinfo {author} {\bibfnamefont {KL}~\bibnamefont {Shaklee}},
  \bibinfo {author} {\bibfnamefont {RF}~\bibnamefont {Leheny}}, \ and\ \bibinfo
  {author} {\bibfnamefont {RB}~\bibnamefont {Zetterstrom}},\ }\bibfield
  {title} {\enquote {\bibinfo {title} {Stimulated {{Emission}} and {{Laser
  Action}} in {{Gallium Nitride}}},}\ }\href {\doibase 10.1063/1.1653730}
  {\bibfield  {journal} {\bibinfo  {journal} {Applied Physics Letters}\
  }\textbf {\bibinfo {volume} {19}},\ \bibinfo {pages} {5} (\bibinfo {year}
  {1971})}\BibitemShut {NoStop}%
\bibitem [{\citenamefont {Suzuki}\ and\ \citenamefont
  {Adachi}(1994)}]{1994Suzuki}%
  \BibitemOpen
  \bibfield  {author} {\bibinfo {author} {\bibfnamefont {Norihiro}\
  \bibnamefont {Suzuki}}\ and\ \bibinfo {author} {\bibfnamefont {Sadao}\
  \bibnamefont {Adachi}},\ }\bibfield  {title} {\enquote {\bibinfo {title}
  {Optical {{Properties}} of {{PbTe}}},}\ }\href {\doibase 10.1143/JJAP.33.193}
  {\bibfield  {journal} {\bibinfo  {journal} {Japanese Journal of Applied
  Physics}\ }\textbf {\bibinfo {volume} {33}},\ \bibinfo {pages} {193}
  (\bibinfo {year} {1994})}\BibitemShut {NoStop}%
\end{thebibliography}%

\end{document}